# On fast charged particles scattering in a thin crystalline undulator

N.F. Shul'ga[a,b,*] and V.I.Truten'[a]

[a] *Akhiezer Institute for Theoretical Physics of National Science Center "Kharkov Institute of Physics and Technology", Akademicheskaya Str.,1, 61108 Kharkov, Ukraine*
[b] *V.N.Karazin Kharkov National University, Svobody Sq. 4, 61022 Kharkov, Ukraine*

**Abstract**

The scattering of fast charged particles in a thin crystalline undulator is considered under conditions when all particles of the beam undergo above-barrier motion with respect to the bent crystalline atomic planes. The consideration is based on the analysis of the particles motion in the continuous potential field of the bent crystal atomic planes and on the simulation of the scattering process taking into account the influence of incoherent effects in scattering on the particles motion. The possibility of manifestation of the effect of multiple volume reflection from bent crystalline atomic planes both for positively and negatively charged particles is shown. The possibility of the beam profiles significant change by means of a small size crystalline undulator is demonstrated.

## 1. Introduction

Upon fast charged particles incidence on a crystal along one of the crystalline planes, the channeling phenomenon is possible, at which the particles perform a finite motion between adjacent crystalline planes (positrons, protons) or in the field of a single crystalline atomic plane (electrons) [1-3]. Such type of particles motion in a crystal can be preserved even at small bending of the crystalline atomic planes. This fact allows the deflecting of the particles beam to an angle coinciding with the crystalline atomic planes bending angle [4, 5]. The case of a crystalline undulator, having periodic bending of crystalline planes, is of special interest.

The particles channeled motion in such undulator leads to a series of interesting peculiarities in radiation characteristics of relativistic electrons and positrons associated with the appearance of additional maxima in their radiation spectrum [6-8] (see also [9] and references therein). However, the motion of channeled particles in the crystalline undulator becomes rather unstable with the increase of the crystalline atomic planes bending amplitude.

The present work is devoted to the study of peculiarities of fast charged particles motion and scattering in a crystalline undulator, in which the amplitude of crystal atomic planes bending exceeds the distance between these planes. It is shown that under these conditions all particles of the beam undergo above-barrier motion with respect to the bent crystalline planes. In this case a significant change of the incident beam shape by the crystals of small thickness is possible.

## 2. Scattering in the field of continuous potential of periodically bent crystal atomic planes

When a fast charged particle moves in a crystal at a small angle to one of the crystalline

---

atomic planes (the $(y,z)$ plane), correlations between its successive collisions with the plane's atoms manifest themselves. These correlations lead to the fact that the particles motion in the crystal in this case is mainly determined by the continuous potential of the crystalline atomic planes $U(x)$, being the crystal atomic potential averaged over the coordinates $(y,z)$ of the plane [1-3]

$$U(x) = \frac{1}{L_y L_z} \int dy dz \sum_n u(\mathbf{r} - \mathbf{r}_n), \tag{1}$$

where $u(\mathbf{r} - \mathbf{r}_n)$- is the potential of a single lattice atom located in the point $\mathbf{r}_n$, $L_y$ and $L_z$ are the crystal sizes along the $y$ and $z$ axes. The summation in (1) is carried out over all the atoms of the crystal.

The continuous potential $U(x)$ is the sum of the continuous potentials $U_p(x)$ of single crystalline planes periodically arranged along the $x$ axis, orthogonal to the $(y,z)$ plane, at distance $d$ one from another:

$$U(x) = \sum_n U_p(x - nd) \tag{2}$$

For a positively charged particle the interplanar potential $U(x)$ is a potential well with the potential distribution close to parabolic. For a negatively charged particle the potential $U(x)$ of a separate crystalline atomic plane is a potential well with the minimum potential value at the point $x=nd$ of the plane position along the $x$ axis. When both positively or negatively charged particles enter the crystal along such crystalline planes, they perform channeled (finite) motion in corresponding potential wells.

In the crystalline undulator the crystalline atomic planes are periodically deformed along the $x$ and $z$ coordinates (see Fig. 1). Assuming, for the sake of simplicity, that the $(y,z)$ plane is deformed according to the harmonic law

$$x(z) = X_0 Sin(\Omega z + \alpha), \tag{3}$$

the continuous potential of a crystalline undulator can be presented in the following form:

$$U(x,z) = \sum_n U_p(x + X_0 Sin(\Omega z + \alpha) - nd), \tag{4}$$

were $\Omega = 2\pi/T$, $T$ is the period of the plane bending along the $z$ axis, $X_0$ is the amplitude of periodic plane bending and $\alpha$ is the bending phase at $z = 0$. It is assumed that $T >> X_0$.

The description of high-energy particles motion in the field of a crystalline undulator can be carried out on the basis of the classical theory. The characteristic values of the particle deflection angles in the field of the crystalline undulator $\theta_{eff}$ relative to the initial direction of the particle beam motion are small compared to unit. In this case, if the particle enters the crystal along the crystalline atomic planes (along the $z$ axis), up to the terms of order $\theta_{eff}^2$ its motion along the $z$ axis will be close to rectilinear one (i.e. $z \approx vt$, where v is the velocity of the particle entering the crystal). With the same accuracy the transverse component of the particle trajectory will be determined by the equation

$$\frac{d^2}{dt^2} x(t) = -\frac{c^2 e}{\varepsilon} \frac{\partial}{\partial x} U(x,z) \bigg|_{z=vt}, \tag{5}$$

where $\varepsilon$ is the particle energy, $e$ is the particle charge and $c$ is the speed of light.

The solution of equation (5) in the field of the crystalline undulator (4) can be found with the use of numerical methods. Fig. 1 shows the typical trajectories of electrons and positrons with the energy $\varepsilon = 10$ GeV in the $(x,z)$ plane of the crystal undulator at the particles entrance into the silicon crystal along the bent crystal planes (110) with the bending amplitude $X_0 = 2d$ and the bending period $T = 10$ μm. (In the considered case the distance between planes is equal to $d = 1.92 \overset{o}{A}$.) Such parameters of the crystalline undulator are close to those which were used in [6]. Different trajectories on Fig. 1 correspond to different points of the particles entrance into the crystal.

In the calculations for positrons the approximation of the continuous potential of an individual silicon crystal plane (110) was used by a function of the form [10]:

$$U_p(x) = \begin{cases} U_0\left(\dfrac{1}{Ch(2\beta x/d)} - \dfrac{1}{Ch(\beta)}\right), & x < d/2 \\ 0, & x > d/2 \end{cases}, \qquad (6)$$

where $U_0$=22,26 eV, $\beta$=4.85. For electrons the potential (6) must be taken with a negative sign.

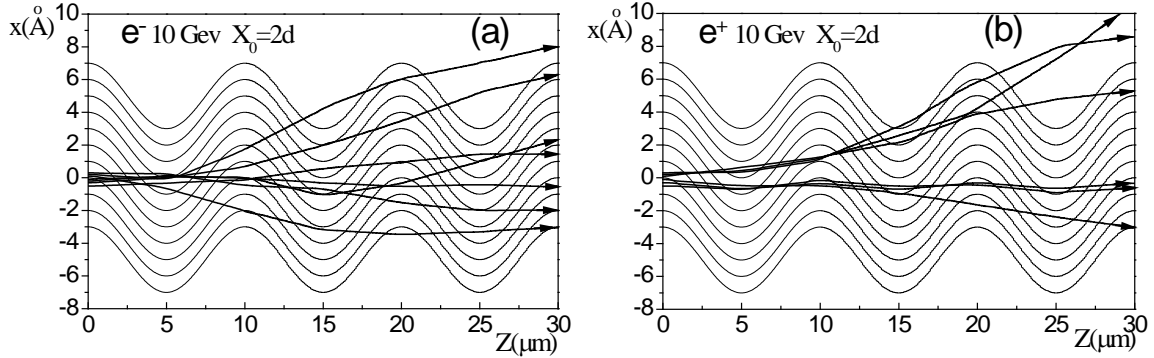

Fig.1. Typical trajectories of negatively (a) and positively (b) charged particles of 10 GeV energy in the (*x,z*) plane of the crystal undulator. The particles enter the silicon crystal along the (110) plane.

The obtained results show that in the case under consideration ($X_0$=2*d*, *T*=10 μm) all beam particles undergo above-barrier motion with respect to the bent crystal atomic planes. The channeling of particles in the crystal is absent in this case. This is due to the fact that the finite motion of fast particles in the crystalline undulator is possible if the angle of maximum inclination of the bent crystal atomic planes $\theta_{und}$ is small in comparison with the critical angle of planar channeling $\theta_c$. The maximum angle of crystalline planes bending $\theta_{und}$, according to (3), is:

$$\theta_{und} = X_0 \Omega, \qquad (7)$$

and in the considered case it is 3.5 times larger than the value of the angle $\theta_c$.

The obtained results also show that in the present case the motion of both the negatively and positively charged particles in the (*x,z*) plane of crystalline undulator is very unstable with respect to a small change of the initial conditions. This is due to the fact that above-barrier particles, when moving in a crystalline undulator, experience multiple intersections of the crystalline atomic planes at different angles. At the same time, practically at each half-period of the (*y,z*) plane bending they get into a regime which is close to the conditions of their volume reflection in a bent crystal [11]. This leads to the fact that the particles, while moving in a crystalline undulator, experience multiple volume reflection at their sliding incidences upon the bent crystalline planes. As a result there occurs a significant broadening of the moving beam in the direction of the crystalline planes bending to the angles of the order of several values of planar channeling critical angle $\theta_c$ (see Fig. 2). Due to this fact, significant changes in the angular characteristics and beam profiles (for example, transformation of a circular beam into a plane-parallel beam) are possible with the help of crystals of small size.

Let us note that in the case under consideration within the first periods of the crystalline atomic planes bending the angular distributions of the particles along the *x* axis are not symmetric with respect to the direction of the incident beam motion. It is associated with the conditions of the beam particles entrance to the bent crystal in the case under consideration, when the beam enters the crystal along the crystalline atomic planes (see Fig. 1). With the increase of the crystal thickness the particle distribution over the scattering angles $\theta_x$ along the $x$ axis becomes more symmetric with respect to the direction of the incident beam motion. (The calculations were carried out for $N = 1000$ particles entering the crystal with different impact parameters $x_0$.)

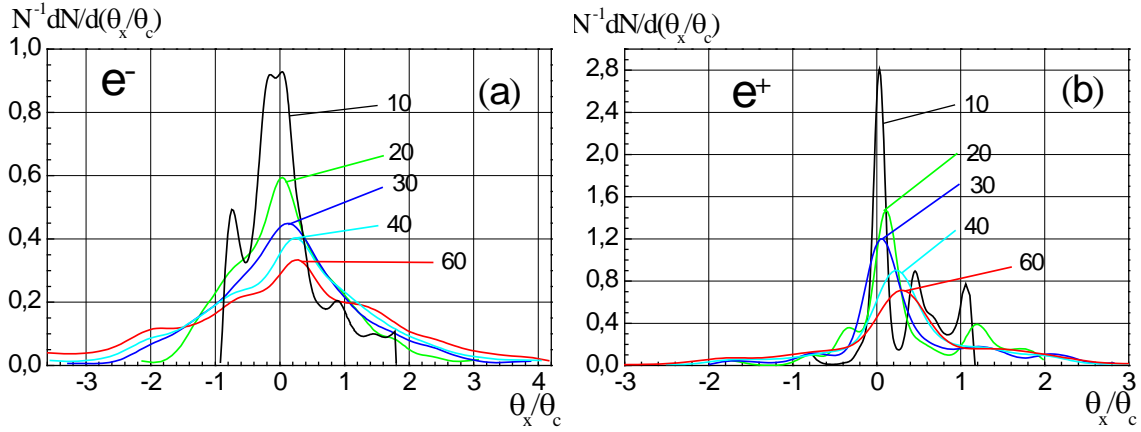

Fig. 2. Vertical profiles of the angular distribution of electrons (a) and positrons (b) of 10 GeV energy for different thicknesses of the crystalline undulator with the bending period $T = 10$ μm. The numbers near the curves correspond to the thickness in μm.

## 3. Accounting for incoherent effects in particles scattering

The results presented in Fig. 1 show that fast particles motion in a crystallized undulator is very complicated even in the simplest case of the continuous potential of bent crystal atomic planes approximation (i.e. without taking into account effects associated with incoherent particle scattering). Taking into account the effects associated with incoherent particle scattering on the heterogeneities of the crystal lattice potential leads to additional instabilities in the particle motion in the crystal. In this case the method of numerical simulation of particle motion in a crystal is of particular importance for the analysis of scattering features under conditions of real particle dynamics in a crystalline undulator.

Fig. 3 and 4 show some simulation results of the angular distributions of negatively and positively charged particles of 10 GeV energy in a crystalline undulator, taking into account incoherent effects in the particles scattering on thermal vibrations of atoms in the crystal. We used in calculations the previously developed program for numerical simulation of the fast charged particle beams passage through a bent crystal [12]. In this program the particle trajectory was calculated by solving step by step the equation of the particle motion in the field of the continuous potential of crystal. The non-coherent effects in scattering are taken into account by playing the value of the particle velocity fluctuation acquired at each step of the trajectory partition due to the lattice potential fluctuations related to the thermal dispersion of the atomic positions in the crystal. For simplicity we proceeded from the simplest assumption that multiple scattering of particles by thermal vibrations of lattice atoms occurs according to the same laws as in amorphous medium, but with the atomic density depending on the distance of the particles from the center of the plane.

The obtained results show that taking into account incoherent effects in the scattering of particles by thermal vibrations of the crystal lattice atoms leads to a small additional broadening of the scattered particle beam perpendicular to the bent crystalline atomic planes along the $x$ axis and to small particle scattering along the bent planes (along the $y$ axis) (see Fig. 3). Incoherent scattering effects lead to the fact that the average values of the scattering angles of the particles along the crystal planes of the atoms are close to the mean scattering angles in the amorphous medium, whereas the scattering in the transverse direction is much more intense than in the amorphous medium.

Accounting for incoherent effects in scattering also leads to the fact that with the increase of the crystal thickness the symmetry of the angular distributions of the scattered particles relative to the direction of motion of the incident beam is established much more rapidly than in the case when incoherent effects in scattering are not taken into account (see Fig.4).

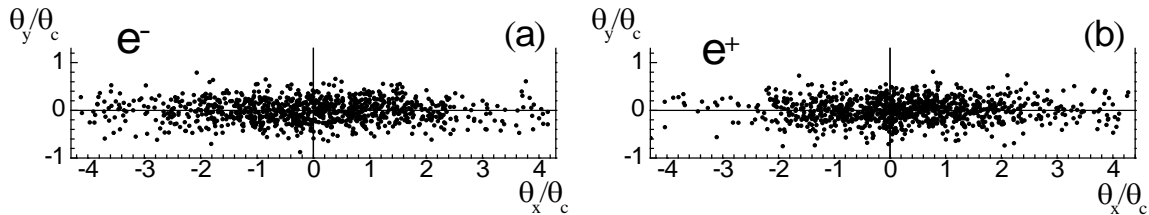

Fig. 3. Angular distribution of electrons (a) and positrons (b) of 10 GeV energy after passing 60 μm crystal undulator.

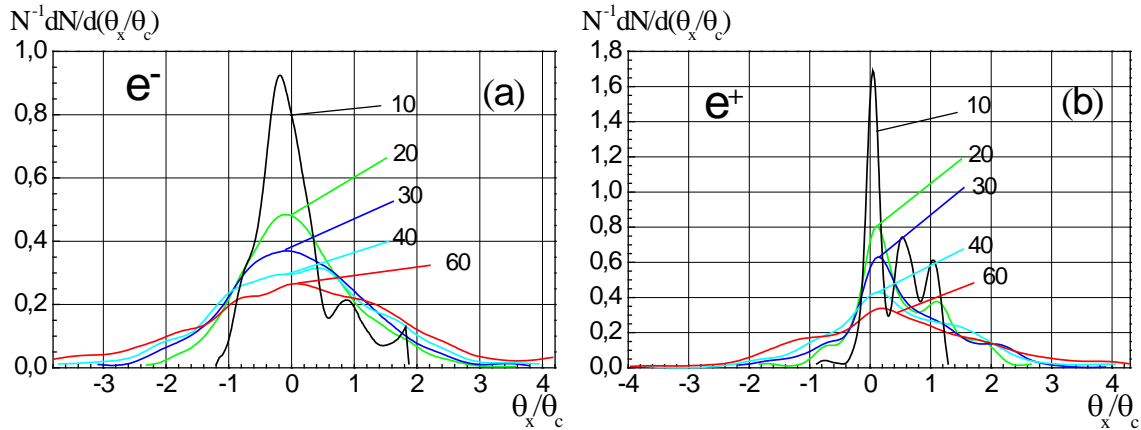

Fig. 4. Vertical profiles of the angular distribution of electrons (a) and positrons (b) of 10 GeV energy for different thicknesses of the crystal undulator. The numbers near the curves correspond to the thickness in μm. The calculations were performed taking into account incoherent effects in the scattering of particles.

## 4. Conclusion

It is shown that when fast charged particles are scattered in a thin crystalline undulator with bending parameters of the crystalline atomic planes, at which the maximum bending angle of the planes $\theta_{und}$ exceeds the value of the critical angle of planar channeling $\theta_c$ and the bending amplitude of the planes of atoms exceeds the interplanar distance, multiple volume reflections of particles from bent crystal atomic planes take place. Due to this fact, a significant change of the incident beam shape by crystals of small thicknesses is possible.

**Acknowledgments**

The work is partly supported by the projects of National Academy of Sciences of Ukraine No. C-2/50-2018 and No. F56-2018.